\documentclass[american,aps,reprint,prl,floatfix]{revtex4-2}
\usepackage[T1]{fontenc}
\usepackage[utf8]{inputenc}
\usepackage{bm}
\usepackage{amsmath}
\usepackage{amssymb}
\usepackage{graphicx}
\usepackage{xcolor}
\usepackage[normalem]{ulem}
\usepackage{newtxtext}
\usepackage{babel}

\begin{document}
\global\long\def\xx{\bm{x}}%
\global\long\def\xo{\xx(0)}%
\global\long\def\tcyc{\tau_{\textrm{p}}}%
\global\long\def\tcoh{\tau_{\textrm{c}}}%
\global\long\def\epcyc{\Sigma_{\tcyc}}%
\global\long\def\ft{t}%
\global\long\def\obs{\phi}%
\global\long\def\Jj{J_{\ft}}%
\global\long\def\im{\operatorname{Im}}
\global\long\def\re{\operatorname{Re}}%
\global\long\def\imL{\im\lambda}
\global\long\def\reL{\re\lambda}%
\global\long\def\epr{\dot{\Sigma}}%
\global\long\def\Corr{C}%
\global\long\def\ii{\mathrm{i}}%
\global\long\def\xt{\xx(\ft)}%
\global\long\def\bb{\omega}%
\global\long\def\St{S_{\ft}}%
\global\long\def\mv{v}%
\global\long\def\flip{s}%
\global\long\def\rate{\gamma}%
\global\long\def\nnn{\eta}%

\newcommand{\PRLsection}[1]{\emph{#1}.---}

\title{Elementary derivation of the dissipation--coherence bound for stochastic oscillators}
\author{Artemy Kolchinsky}
\affiliation{ICREA-Complex Systems Lab, Universitat Pompeu Fabra, 08003 Barcelona, Spain}
\affiliation{Universal Biology Institute, The University of Tokyo, 7-3-1 Hongo, Bunkyo-ku, Tokyo 113-0033, Japan}
\email{artemyk@gmail.com}

\begin{abstract}
The dissipation--coherence bound is a conjectured tradeoff between entropy production and the quality of stochastic oscillations. We show that this bound can be derived by combining the higher-order ``thermodynamic uncertainty relation'' with a simple condition on phase-current fluctuations. In one-dimensional cyclic systems, our proposed condition is shown to be equivalent to the dissipation--coherence bound itself. Our approach yields an elementary proof in the weak-noise Gaussian regime and extends naturally to some non-Gaussian systems, as we illustrate with a run-and-tumble particle. Finally, we contrast current-based and spectral formulations of the dissipation--coherence bound.
\end{abstract}
\maketitle

\PRLsection{Introduction} Sustaining coherent oscillations in an autonomous overdamped system carries an unavoidable thermodynamic cost. This idea was first developed in the work of Gaspard~\citep{gaspard2002correlation}, and it has since inspired a line of research on the free-energy cost of biochemical clocks and other nonequilibrium oscillators~\citep{cao_free-energy_2015,barato_coherence_2017,marsland_thermodynamic_2019,pearson_measuring_2021,remlein2022coherence,ohga_thermodynamic_2023,shiraishi_entropy_2023,kolchinsky_thermodynamic_2024,xu_thermodynamic_2025,zheng_trade-off_2025,kolchinsky2026cycleaffinitywindinglocalize,gu_dissipation-coherence_2026,nagayama2026oscillatory}. Other research has also investigated the thermodynamic cost of coherent oscillations
in underdamped~\citep{pietzonka_classical_2022,gopal_thermodynamic_2024},
periodically driven~\citep{barato_cost_2016,koyuk_operationally_2019},
and quantum~\citep{erker_autonomous_2017,woods_quantum_2022,meier_fundamental_2023,meier_precision_2025} systems,
finding qualitatively different tradeoffs in each case. 

The tradeoff between oscillations and thermodynamic cost is sometimes termed the \emph{dissipation--coherence bound} (DCB). The DCB is a conjectured inequality which states that, in autonomous overdamped systems, the entropy production (EP) per cycle obeys~\citep{santolin2025dissipation,nagayama2025}
\begin{equation}
\epcyc\ge4\pi^{2}\frac{\tcoh}{\tcyc} \,,
\label{eq:res}
\end{equation}
where $\tcyc$ is the period of the oscillator and $\tcoh$ is its correlation time. Equivalently, this bound may be written using the entropy production rate (EPR) $\epr=\epcyc/\tcyc$ and oscillator frequency $\omega=2\pi/\tcyc$ as
\begin{equation}
\epr\ge\omega^{2}\tcoh.
\label{eq:dc2}
\end{equation}

The dissipation--coherence bound is expected to hold in a broad class of systems. To date, however, it has only been proven rigorously for stochastic limit cycles in the weak-noise regime~\citep{santolin2025dissipation,nagayama2025}. The original proof by Santolin and Falasco~\citep{santolin2025dissipation} employed sophisticated analytical techniques
and technical assumptions (an identity-like diffusion matrix or proximity to a Hopf bifurcation).
Nagayama and Ito~\citep{nagayama2025} proved the same result without
these additional assumptions, while also highlighting connections
to the short-time thermodynamic uncertainty relation (TUR) and thermodynamic
speed limits. %

In this paper, we show that the DCB can be easily derived by combining the higher-order TUR with a simple condition on phase-current fluctuations. We first use the TUR to bound entropy production in terms of oscillator frequency and phase diffusion. Then, the proposed condition implies the DCB by relating the phase-diffusion coefficient to the correlation time. Our condition can be analytically verified in many systems of interest. Moreover, in one-dimensional systems the condition is both sufficient and necessary, hence equivalent to the DCB. %

Our analysis provides an elementary proof of the DCB for oscillators with Gaussian fluctuations of the phase current, including, as a special case, stochastic limit cycles in the weak-noise regime. It also provides a new route for proving the bound in some systems with non-Gaussian fluctuations, as we demonstrate on a model of a simple run-and-tumble particle.

\PRLsection{Setup} We consider a stationary overdamped Markovian system with a fluctuating state $\xx$. The system is also equipped with a phase function $\theta(\xx)\in[0,2\pi)$ that maps each state to a phase. 
Given a trajectory $\{\xx(s):s\in[0,\ft]\}$, the \emph{phase current} is defined as
\begin{equation}
\Jj=\int_{0}^{\ft}\nabla \theta \circ d\xx =\Theta(\ft)-\Theta(0),
\label{eq:jt}
\end{equation}
where $\Theta$ is the unwrapped phase, so $\theta(\xx(s))=\Theta(s)\bmod 2\pi$. 

For concreteness, one may imagine an overdamped Langevin system with state $\xx(t)\in\mathbb{R}^{d}$. In this case, the phase function $\theta$ should be twice continuously differentiable (up to branch cuts) on the support of the steady-state distribution, so that the stochastic integral is well-defined. As long as that smoothness condition is satisfied, the phase function may be defined in various ways. For instance, it may be defined by projecting the state onto the tangential direction of a weak-noise limit cycle~\citep{santolin2025dissipation} or by using the phase angle of the second eigenfunction of the backward Fokker--Planck operator~\citep{thomas2014asymptotic}.

Alternatively, one may consider a discrete system described by a Markov jump process. After associating each discrete state with a phase, the phase current $\Jj$ may be defined as the sum of the unwrapped phase increments across jumps.

We define the oscillator frequency as
\begin{align}
\omega= \langle \Jj\rangle/\ft\,,
\end{align}
where $\langle \cdot \rangle$ indicates stationary expectation. By stationarity, this definition is independent of $t$. In addition, we introduce the complex phase observable $\obs(\xx)=e^{-\ii\theta(\xx)}$. Importantly, the autocorrelation function of $\obs(\xx)$ can be expressed as the characteristic function of $\Jj$,
\begin{align}
\Corr_{t}
&:=\langle \obs(\xt)^{*}\obs(\xo)\rangle \nonumber\\
&=\langle e^{\ii[\theta(\xt)-\theta(\xo)]}\rangle=\langle e^{\ii\Jj}\rangle.
\label{eq:cc}
\end{align}
The last equality follows since $\Jj=\Theta(\ft)-\Theta(0)$ differs from $\theta(\xt)-\theta(\xo)$ by an integer multiple of $2\pi$. 
 We define the correlation time $\tcoh$ by
\begin{equation}
\frac{1}{\tcoh}
:=-\limsup_{\ft\to\infty}\frac{1}{t}\ln\left|\Corr_{t}-\Corr_{\infty}\right|
\label{eq:ac}\,,
\end{equation} 
assuming the limit $\Corr_\infty:=\lim_{t\to\infty}\Corr_t$ exists. In systems with mixing dynamics, $\Corr_\infty=\vert \langle\obs\rangle \vert^2$.

In the following, we will refer to the cumulant generating function (CGF) of the phase current,
\begin{align}
\psi_{t}(\alpha):=\ln\langle e^{\alpha \Jj}\rangle =\sum_{n=1}^{\infty}\frac{\alpha^n}{n!}\kappa_t^{(n)}\,,%
\label{eq:psi}
\end{align}
where $\kappa_t^{(n)}:=\partial^{n}_\alpha \psi_t\vert_{\alpha=0}$ is the $n$th cumulant.%

\PRLsection{Dissipation--coherence bound from the higher-order TUR} We now demonstrate that the DCB~\eqref{eq:dc2} follows by combining the TUR and a simple condition on the phase-current fluctuations.

We employ the higher-order TUR proposed by Dechant and Sasa~\citep{dechant2020fri}, 
\begin{equation}
\epr\ge\frac{\omega^{2}\alpha^{2}}{\psi_{t}(\alpha)/t-\alpha\omega}\,,
\label{eq:hotur}
\end{equation}
where $\psi_{t}$ is the CGF from Eq.~\eqref{eq:psi}. 
This bound holds for all timescales $t \ge 0$ and counting fields $\alpha\in \mathbb{R}$ (with $t=0$ and $\alpha=0$ treated by continuity). 
Optimizing over $t$ and $\alpha$ gives the tightest bound
\begin{equation}
\epr\ge\frac{\omega^{2}}{\mathcal{D}_{*}}\,,
\label{eq:hoturopt}
\end{equation}
where we introduced
\begin{equation}
\mathcal{D}_{*}:=\inf_{t> 0,\alpha\in\mathbb{R}\setminus\{0\}}
\frac{\psi_{t}(\alpha)/t-\alpha\omega}{\alpha^{2}}\,.
\label{eq:dstar}
\end{equation}
The quantity $\mathcal{D}_{*}$ can be interpreted as a generalized phase-diffusion coefficient.

We compare the conjectured DCB~\eqref{eq:dc2}, involving the correlation time $\tcoh$, with the optimized higher-order TUR~\eqref{eq:hoturopt}, involving the generalized diffusion coefficient $\mathcal{D}_*$. The former follows from the latter whenever
\begin{equation}
\frac 1 \tcoh \ge \mathcal{D}_{*} \,.
\label{eq:dstarcrit}
\end{equation}
Inspired by this, we propose inequality~\eqref{eq:dstarcrit} as a simple sufficient condition for the DCB~\eqref{eq:dc2}. The advantage of this approach is that condition~\eqref{eq:dstarcrit} can be analytically verified in many systems of interest, since it involves two mathematically related objects: the characteristic function and the cumulant generating function.

In general, our proposed condition is more restrictive than the DCB, as the latter may sometimes hold even if the former does not. 
It may be possible to derive more general sufficient conditions by using more sophisticated TURs. For example, Dechant and Sasa derived a refinement of their higher-order TUR involving continuous time-reversal~\citep{dechant2021continuous}, which provides a more general sufficient condition at the cost of added complexity. Conversely, simpler but less general conditions follow by considering simpler TURs. For example, we may recall the standard finite-time TUR~\cite{pietzonka2017finite,horowitz2017proof}, 
\begin{align}
\epr\ge\frac{\omega^{2}}{D_{t}}\,,
\end{align}
where we introduced the time-$t$ phase-diffusion coefficient:
\begin{align}D_t :=\frac{\mathrm{Var}(\Jj)}{2t}\equiv \frac{\kappa_t^{(2)}}{2t}\equiv \lim_{\alpha\to 0}
\frac{\psi_{t}(\alpha)/t-\alpha\omega}{\alpha^{2}}\,.
\end{align}
Since this holds for all $t$, it implies 
the DCB whenever
\begin{equation}
\frac 1 \tcoh \ge {\inf_{t > 0}D_{t}}\,.
\label{eq:d} 
\end{equation}
Condition~\eqref{eq:d} is often more useful than~\eqref{eq:dstarcrit}, since the finite-time phase-diffusion coefficient is typically easier to compute or measure than the full CGF. 
However, condition~\eqref{eq:d} is more restrictive than~\eqref{eq:dstarcrit}, since $\mathcal{D}_* \le D_t$ for all $t$.

\PRLsection{One-dimensional systems} 
As mentioned, our condition~\eqref{eq:dstarcrit} is generally only sufficient, not necessary, for the DCB~\eqref{eq:dc2}. However, in one-dimensional systems --- such as overdamped diffusion on a ring and unicyclic Markov chains --- our condition is both sufficient and necessary, and the two inequalities~\eqref{eq:dc2} and~\eqref{eq:dstarcrit} are equivalent. To demonstrate this equivalence, we show that~\eqref{eq:dstarcrit} follows from~\eqref{eq:dc2}.

Given a one-dimensional system, let the random variable $N_{t}$ denote the net number of clockwise cycles that the system has completed by time $t$. In addition, let the random variable $\sigma_t$ denote the trajectory EP the system has accumulated by time $t$, so that $\epr = \langle \sigma_t\rangle/t$. The phase current and trajectory EP have the asymptotic form
\begin{equation}
\Jj=2\pi k N_{t} +O(1),
\qquad
\sigma_t=\mathcal{A} N_{t}+O(1),
\label{eq:1dpath}
\end{equation}
where $k$ is the number of oscillator cycles per state cycle and $\mathcal{A}$ is the cycle affinity. Importantly, the phase current is asymptotically proportional to the stochastic EP: $\alpha_{*}\Jj=-\sigma_t+O(1)$ for $\alpha_{*}=-\mathcal{A}/(2\pi k)$. We plug this $\alpha_*$ into Eq.~\eqref{eq:dstar} to bound $\mathcal{D}_*$ as 
\begin{equation}
\mathcal{D}_{*}\le\lim_{t\to \infty}
\frac{\psi_{t}(\alpha_*)/t-\alpha_*\omega}{\alpha_*^{2}}=\frac{-\omega}{\alpha_*}=\frac{\omega^{2}}{\epr}\,.
\label{eq:dstarle}
\end{equation}
Here, we first applied the variational definition~\eqref{eq:dstar} of $\mathcal{D}_*$, then used $\lim_{t\to \infty}\psi_{t}(\alpha_{*})/t= 0$, which follows from the integral fluctuation theorem $\langle e^{-\sigma_t}\rangle=1$, and finally substituted $\epr=\mathcal{A} \omega/(2\pi k)$. Inequality~\eqref{eq:dstarle} is the reverse of inequality~\eqref{eq:hoturopt}, implying that this TUR is saturated in one-dimensional systems. 
Finally, we assume that the DCB holds, so $\epr\ge\omega^{2}\tcoh$. Combining with $\mathcal{D}_{*}\le {\omega^{2}}/{\epr}$ and rearranging gives condition~\eqref{eq:dstarcrit}.

\PRLsection{Gaussian fluctuations and weak-noise regime} We now show that our analysis provides 
an elementary derivation of the DCB in the case of Gaussian fluctuations. 

We consider the case where the phase current is ``{asymptotically Gaussian}''. By this, we mean that the autocorrelation function obeys 
\begin{equation}
\vert \Corr_{t}\vert \equiv \vert \langle e^{\ii\Jj}\rangle \vert=e^{-tD_{t}+o(t)}\,.
\label{eq:gg}
\end{equation}
When the fluctuations are exactly Gaussian, $C_t=\langle e^{\ii\Jj}\rangle=e^{\ii t\omega-tD_{t}}$, so this form holds without any sublinear correction. 
Eq.~\eqref{eq:gg} implies that $\Corr_\infty=0$ whenever $D_\infty:=\lim_{t\to\infty}D_t>0$. 
Plugging into Eq.~\eqref{eq:ac} and taking limits then gives
\begin{equation}
\frac{1}{\tcoh}=D_{\infty}\,.
\label{eq:gausscrit}
\end{equation}
Thus, oscillators with asymptotically Gaussian fluctuations satisfy the simpler criterion~\eqref{eq:d} and thus obey the DCB.

For stochastic oscillators in the weak-noise regime, the phase current can be shown to be Gaussian (see Refs.~\cite[Eq.~(54)]{gaspard2002correlation} and \cite[SM]{santolin2025dissipation}). In this way, our approach provides an elementary proof of the DCB in the weak-noise regime, without requiring the sophisticated technical machinery used in earlier derivations~\citep{santolin2025dissipation,nagayama2025}.

\PRLsection{Example: run-and-tumble particle} We illustrate our approach on a system with non-Gaussian fluctuations: a run-and-tumble particle on a ring. The system is described by a phase variable $\theta(t)\in[0,2\pi)$ and an internal state $\flip(t)\in\{+1,-1\}$ that flips at rate $\rate$. %
The phase variable evolves according to a Langevin equation,
\begin{equation}
d\theta(t)=(\bb+\mv\,\flip(t))\,dt+\sqrt{2/\beta}\,dW_{t} \,,
\label{eq:tele_sde}
\end{equation} 
where $\bb$ is the average frequency, $\mv$ is the amplitude of frequency fluctuations, and $\beta$ is the inverse temperature of a coupled bath. Related variants of this model were studied in Refs.~\cite{malakar2018steady,bao2023improving}. 
To compute the steady-state EPR, we use the fact that, when conditioned on the internal state $\flip(t)=\pm 1$, the dynamics is a simple drift-diffusion with velocity $\bb \pm \mv$ and inverse temperature $\beta$, so the conditional EPR is $\beta(\bb\pm\mv)^2$. Averaging over the two equiprobable internal states $\flip(t)$ gives 
\begin{align}
\epr=\frac{\beta}{2}\big[(\bb+\mv)^2+(\bb-\mv)^{2} \big]=\beta(\bb^{2}+\mv^{2})\,.
\end{align}
Flips of the internal state $\flip(t)$ satisfy detailed balance and therefore do not contribute to steady-state EP. 

The phase current is
\begin{equation}
\Jj=\int_{0}^{\ft}d\theta(s)=\ft\bb+\mv\St+\sqrt{2/\beta}\,W_{\ft},
\label{eq:tele_J}
\end{equation}
where $W_t$ is a Wiener process and $\St:=\int_{0}^{\ft}\flip(u)\,du$ is the integrated telegraph process. 
In the following, we calculate the correlation time and the CGF using the identity
\begin{align}
\langle e^{\alpha \Jj}\rangle&=e^{\alpha \ft \bb}\langle e^{\alpha\mv \St}\rangle\langle e^{\alpha \sqrt{2/\beta} W_t}\rangle \nonumber \\
&=e^{\alpha t\bb+ \alpha^2 t/\beta}\langle e^{\alpha\mv\St}\rangle\,.%
\label{eq:genfunc00}
\end{align}
Here, we first used that $W_t$ and $\St$ are independent, then employed the identity $\langle e^{\alpha\sqrt{2/\beta}W_t}\rangle=e^{\alpha^2 t/\beta}$. The generating function of the integrated telegraph process is~\citep[Thm.~2.4]{kolesnik_telegraph_2013},
\begin{align}
\langle e^{\alpha\mv\St}\rangle=
e^{-\rate\ft}\big\{\cosh [\ft \eta(\alpha)]+ \frac{\rate}{\eta(\alpha)}\sinh [\ft \eta(\alpha)]\big\},
\label{eq:genfunc01}
\end{align}
where for convenience we introduced $\eta(\alpha):=\sqrt{\rate^2+ \alpha^2\mv^2}$. 

After some rearranging, we may use these identities to express the 
autocorrelation function $\Corr_{t}=\langle e^{\ii\Jj}\rangle$ as
\begin{align}
\Corr_t
=
e^{(-1/\beta-\gamma)t}
\Big[
\frac{\nnn+\gamma}{2\nnn} e^{(\ii\bb + \nnn) t}
+
\frac{\nnn-\gamma}{2\nnn} e^{(\ii\bb - \nnn) t}
\Big]\,,
\label{eq:CorrTRTP}
\end{align}
where $\nnn \equiv \eta(\ii) = \sqrt{\rate^2-\mv^2}$. %
Plugging into Eq.~\eqref{eq:ac} and taking limits gives the correlation time,
\begin{equation}
\frac{1}{\tcoh}=\frac{1}{\beta}+\rate-\re\nnn \,.
\label{eq:telecoh}
\end{equation}
The autocorrelation function exhibits a regime change at $\mv=\rate$. When $\mv< \rate$, $\nnn$ is real and $\Corr_t$ is dominated by a single damped oscillation at frequency $\bb$ and decay rate $1/\beta +\rate -\nnn$. When $\mv >\rate$, $\nnn$ is imaginary and $\Corr_t$ has two oscillatory contributions with frequencies $\bb \pm \im \nnn$ and the same decay rate $1/\beta + \rate$.

We calculate the long-time scaled CGF using the same identities, %
\begin{align}
\lim_{t\to\infty}\frac{1}{t}\ln\langle e^{\alpha\Jj}\rangle
=\alpha \bb+\frac{\alpha^{2}}{\beta}-\rate+\eta(\alpha)\,.%
\end{align}
By considering the higher scaled cumulants, we can verify that 
the long-time statistics of the phase current are non-Gaussian for all $\mv>0$. For instance, the fourth scaled cumulant can be calculated as $\lim_{t\to\infty}\kappa_{t}^{(4)}/t=-3\mv^{4}/\rate^{3}\neq0$. Moreover, for real $\alpha$, the scaled CGF is smooth as a function of $\mv$. Thus, unlike the correlation time $\tcoh$, the scaled cumulants do not exhibit any singularity at $\mv=\rate$.

\begin{figure}[t]
\includegraphics[width=1\columnwidth]{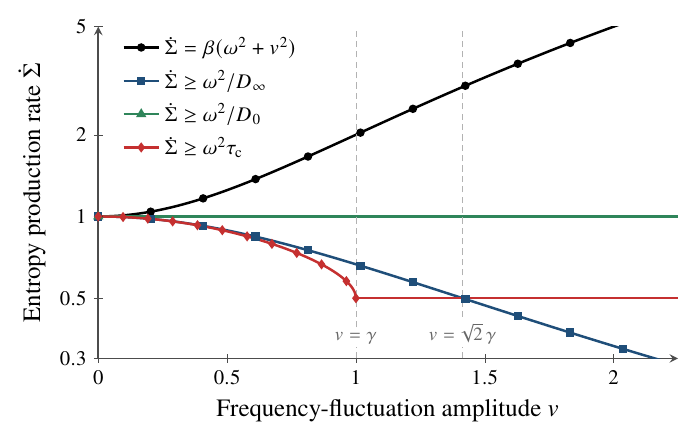}
\caption{\textbf{Run-and-tumble particle on a ring}. %
We plot the EPR $\dot{\Sigma}$, the long-time TUR $\dot{\Sigma}\ge \bb^2/D_\infty$, the short-time TUR $\dot{\Sigma} \ge \bb^2/D_0$, and the DCB $\dot{\Sigma}\ge \bb^2 \tcoh$ as a function of the frequency-fluctuation amplitude $\protect\mv$ (other parameters: $\beta=\protect\rate=\protect\bb=1$). The two vertical lines mark the thresholds $\protect\mv = \protect\rate$, where the autocorrelation function splits into two oscillatory modes, and $\protect\mv=\sqrt{2}\protect\rate$, where the DCB becomes tighter than the long-time TUR.}
\label{fig:rtp}
\end{figure}

Finally, we derive the DCB %
using the simple diffusion-based condition~\eqref{eq:d}.
The time-$t$ diffusion coefficient is
\begin{equation}
D_{t}=\frac{1}{\beta}+\frac{\mv^{2}}{2\rate}-\frac{\mv^{2}}{4\rate^{2}t}(1-e^{-2\rate t}).
\label{eq:teleDt}
\end{equation}
Since $D_{t}$ is monotone increasing, its minimum is given by the short-time diffusion coefficient, $D_{0}:=\inf_{t > 0}D_{t}=1/\beta$.
Using Eq.~\eqref{eq:telecoh}, we verify that $1/\tcoh\ge D_{0}$ always. Therefore, condition~\eqref{eq:d} is satisfied and the DCB $\epr\ge\bb^{2}\tcoh$ holds. 

The inequality $1/\tcoh\ge D_{0}$ implies that, in this model, the DCB is weaker than the short-time TUR $\epr\ge\bb^{2}/D_0$. We may also compare with the long-time TUR $\epr\ge\bb^{2}/D_\infty$, where $D_{\infty}=1/{\beta}+{\mv^{2}}/{2\rate}$. 
Comparing to Eq.~\eqref{eq:telecoh} shows that the DCB is stronger than the long-time TUR $\epr\ge\bb^{2}/D_\infty$ whenever $\mv>\sqrt{2}\rate$. 

In Fig.~\ref{fig:rtp}, we plot the EPR $\dot{\Sigma}$, the long-time TUR, the short-time TUR, and the DCB as a function of the frequency-fluctuation amplitude $\mv$ (we set $\beta=\rate=\bb=1$). The EPR increases with $\mv$, while the TURs and the DCB remain constant or decrease. All bounds coincide in the Gaussian limit $\mv\to0$.

\PRLsection{Relation to spectral bound} 
Above, we considered the DCB where the frequency $\omega$ and correlation time $\tcoh$ are defined in terms of the phase current $\Jj$. However, there is also a spectral variant of the DCB, where the frequency and correlation time are defined in terms of the eigenvalue associated with the slowest relaxation mode. 

Specifically, Oberreiter~\emph{et al.}~\citep{oberreiter2022universal} conjectured that Markovian systems obey the spectral inequality
\begin{align}
\epr\ge\frac{(\imL)^{2}}{-\reL}\qquad\text{whenever}\qquad \vert\imL\vert\ge -\reL\,,
\label{eq:ob}
\end{align}
where $\lambda$ is the slowest nonzero eigenvalue of the generator. 
Eq.~\eqref{eq:ob} is analogous to the DCB~\eqref{eq:dc2}, except that the oscillator frequency and correlation time are defined in terms of the eigenvalue, as $\vert\imL\vert$ and $-1/\reL>0$, instead of the phase current $\Jj$. Although inequality~\eqref{eq:ob} is supported by extensive numerical evidence, it remains unproven. (A weaker version was recently shown in Ref.~\cite{gu_dissipation-coherence_2026}.) 

Here, we compare the current-based and spectral definitions of the correlation time and oscillator frequency. We focus on the generic case where the autocorrelation $\Corr_t$ of the phase observable 
$\obs(\xx)=e^{-\ii\theta(\xx)}$ is asymptotically dominated by a single eigenmode. 
By this, we mean that the autocorrelation function at large $t$ can be expressed as
\begin{equation}
\Corr_{t}=\langle \obs(\xt)^{*}\obs(\xo)\rangle
= \Corr_{\infty}+ e^{t\lambda+ o(t)}\,,
\label{eq:diag}
\end{equation}
where $o(t)$ represents a sublinear correction that is continuous in $t$. 
This expression can be derived by a Jordan decomposition (for nondiagonalizable operators, polynomial prefactors are absorbed into the sublinear term). Here, $\lambda$ refers to the slowest nonzero eigenvalue that contributes to $\Corr_t$ with nonzero coefficient and, if both members of a conjugate pair contribute, it refers to the member whose coefficient is strictly larger in magnitude. In generic cases, $\lambda$ is the same eigenvalue that appears in the spectral DCB~\eqref{eq:ob}, though the two may differ if $\obs$ is orthogonal to the slowest mode of the generator.

We now compare the current-based definition of the correlation time with the spectral one. 
To do so, we plug Eq.~\eqref{eq:diag} into Eq.~\eqref{eq:ac} and take limits to give ${\tcoh}=-1/\reL$. Therefore, as long as Eq.~\eqref{eq:diag} holds, the two definitions of the correlation time are equivalent.

On the other hand, the current-based definition of the frequency does not necessarily agree with the spectral one. 
Nonetheless, we show that the definitions coincide when two 
conditions are satisfied, in addition to Eq.~\eqref{eq:diag} being valid. %
First, the autocorrelation vanishes at long times, $\Corr_\infty=0$. 
Second, phase-current fluctuations are symmetric at long times, meaning the distribution of $\Jj-\langle \Jj\rangle=\Jj-\omega t$ is invariant under sign reversal. 

Combining the condition $\Corr_\infty =0$ with Eq.~\eqref{eq:diag} implies that $\Corr_t\neq0$ at long times. 
This permits us to continuously unwrap the autocorrelation phase as 
\[\im\ln\Corr_{t} = t\imL+o(t).\]
At the same time, we can write $\Corr_t=\langle e^{\ii \Jj }\rangle =e^{\ii\omega t}\langle e^{\ii(\Jj-\omega t)}\rangle$. %
The last factor is continuous in $t$ and, due to the symmetry condition, real-valued. Since $\Corr_t$ is eventually nonzero, this factor must have a fixed sign at large $t$, therefore
\[
\im\ln \Corr_t=\omega t+O(1).
\]
Combining and taking limits gives $\omega = \imL$, so the current-based and spectral definitions of the frequency coincide. Consequently, whenever $\obs$ couples to the slowest mode of the generator, the current-based DCB~\eqref{eq:dc2} and the spectral DCB~\eqref{eq:ob} are equivalent (up to the validity condition $\vert\imL\vert\ge -\reL$). 

Oscillators with exactly Gaussian phase fluctuations have vanishing long-time autocorrelation and symmetric fluctuations. In such systems, as long as Eq.~\eqref{eq:diag} holds, the current-based and spectral definitions of the frequency are equivalent. %

In the run-and-tumble model, the autocorrelation function~\eqref{eq:CorrTRTP} vanishes at long times and $\Jj-\bb\ft=\mv\St+\sqrt{2/\beta}W_{\ft}$ has symmetric fluctuations, since both $\St$ and $W_{\ft}$ are symmetric. 
However, this model provides an interesting example where Eq.~\eqref{eq:diag} may no longer be valid, in which case the equivalence between current-based and spectral definitions breaks down. %
We consider two regimes. When $\mv<\rate$, $\Corr_t$ is dominated by a single oscillation at frequency $\bb$ and decay rate $1/\beta+\rate-\nnn$. Eq.~\eqref{eq:diag} is valid, so 
the current-based and spectral definitions of the frequency agree. In contrast, when $\mv>\rate$, the autocorrelation function in Eq.~\eqref{eq:CorrTRTP} contains two oscillatory terms at different frequencies
$\bb\pm\im\nnn$ and the same decay rate
$1/\beta+\rate$. In this regime, the slowest oscillatory mode is no longer unique and Eq.~\eqref{eq:diag} is no longer valid. 
This shows that the current-based definitions of correlation time and frequency are more general than the spectral ones, since they remain well-defined even when phase-current autocorrelations are not dominated by a single oscillatory mode.

\vspace{5pt}

\emph{Acknowledgments}.--- I thank Ryuna Nagayama and Sosuke Ito for helpful comments and suggestions. A. K. is partly supported by the John Templeton Foundation (Grant No. 62828) and by the European Union's Horizon 2020 research and innovation programme under the Marie Sk\l{}odowska-Curie Grant Agreement No. 101068029.

\bibliography{refs}

\end{document}